\documentclass[preprintnumbers,twocolumn,showpacs,amsmath,amssymb]{revtex4}

\usepackage{epsfig}
\usepackage{graphicx}
\usepackage{dcolumn}
\usepackage{bm}

\newcommand{\be}{\begin{equation}}
\newcommand{\ee}{\end{equation}}

\newcommand{\bea}{\begin{eqnarray}}
\newcommand{\eea}{\end{eqnarray}}
\newcommand{\bfig}{\begin{figure}}
\newcommand{\efig}{\end{figure}}
\newcommand{\bc}{\begin{center}}
\newcommand{\ec}{\end{center}}
\def\ycut{y_{{\rm cut}}}

\begin{document}
\preprint{ZU-TH 03/08, IPPP/08/05}

\title{Jet rates in electron-positron annihilation at  
${\cal O}(\alpha_s^3)$ in QCD}

\author{A.\ Gehrmann-De Ridder$^a$, T.\ Gehrmann$^b$, E.W.N.\ Glover$^c$,
G.\ Heinrich$^c$}
 \affiliation{$^a$ Institute for Theoretical Physics, ETH, CH-8093 Z\"urich,
Switzerland\\
$^b$ Institut f\"ur Theoretische Physik,
Universit\"at Z\"urich, CH-8057 Z\"urich, Switzerland\\
$^c$ Institute of Particle Physics Phenomenology, 
        Department of Physics,
        University of Durham, Durham, DH1 3LE, UK}

\date{\today}

\begin{abstract}
We compute  production rates for two, three, four and five jets in electron-positron annihilation at the third order in the QCD coupling constant. At this order, three-jet production is described to next-to-next-to-leading order (NNLO) in perturbation theory while the two-jet rate is obtained at next-to-next-to-next-to-leading order (N$^3$LO). Our results yield an improved perturbative description of the dependence of jet multiplicity on the jet resolution parameter, $\ycut$, particularly at small values of $\ycut$. 
\end{abstract}

\maketitle


Jet observables in  electron--positron
annihilation play an outstanding role in 
studying the dynamics of 
the strong interactions, 
described by the theory of quantum chromodynamics (QCD,~\cite{qcd}).  
The initial experimental observation
of three-jet events at PETRA~\cite{petra}, in agreement with the theoretical
prediction~\cite{ellis}, provided  first evidence for the gluon, and thus 
strong initial 
support for the correctness of QCD. Subsequently 
the three-jet rate  and related event shape observables were used
for the precise determination  of the QCD coupling constant $\alpha_s$
(see~\cite{reviews,bethke} for a review), and four-jet observables helped 
substantially to confirm the gauge group structure of QCD by 
firmly establishing the gluon self-coupling~\cite{expgauge}.

Jets are defined using 
a jet algorithm, which describes how to recombine the momenta 
of all hadrons in an event to 
form the jets. A jet algorithm consists of  two ingredients: a 
distance measure and a recombination procedure.
The distance measure 
is computed for each pair of momenta to select the pair with the smallest 
separation. This pair of momenta then is combined 
according to the recombination procedure into a joint momentum, 
if its separation is below a pre-defined resolution parameter $\ycut$.
Improving upon the JADE algorithm~\cite{jade}, which uses the pair
invariant mass as distance measure, several jet algorithms 
have been proposed for $e^+e^-$ collisions: Durham~\cite{durham},
Geneva~\cite{geneva} and Cambridge~\cite{cambridge}. Among those, 
the Durham algorithm has been the most widely used by experiments at 
LEP~\cite{aleph,opal,delphi,l3} and SLD~\cite{sld}, as well as 
in the reanalysis of earlier data at lower energies from 
JADE~\cite{jadenew}.

The Durham jet algorithm clusters particles into jets 
by computing the distance measure
\begin{equation}
y_{ij,D} = \frac{2 \, {\rm min} (E_i^2,E_j^2) (1-\cos \theta_{ij})}
{E_{{\rm vis}}^2}
\end{equation}
for each pair ($i,j$) of particles, $E_{{\rm vis}}$ 
denotes the energy sum of all particles in the final state.
The pair with the lowest 
$y_{ij,D}$ is replaced by a pseudo-particle whose four-momentum is 
given by the 
sum of the four-momenta of particles $i$ and $j$ ('E' recombination 
scheme). This procedure is repeated as long as pairs with invariant 
mass below the predefined resolution parameter
$y_{ij,D}<\ycut$  are found. Once the clustering is terminated, the 
remaining (pseudo-)particles are the jets. It is evident that a large 
value of $\ycut$ will ultimately result in 
the clustering all particles 
into only two jets, while higher jet multiplicities will become more and 
more frequent as $\ycut$ is lowered. In experimental jet measurements, 
one therefore  studies the jet rates (jet cross sections normalized to 
the total hadronic cross section) as function of the jet resolution 
parameter $\ycut$.

The theoretical prediction of jet cross sections 
is made within perturbative QCD, where the same jet algorithm is applied 
to the momenta of final state partons. The QCD description of jet production 
is either based on a fixed-order calculation, which uses exact parton-level
matrix elements (including higher order corrections if available)
for a given jet multiplicity, or by a parton shower, which 
is based on the leading-order matrix element for two-jet production only, 
and generates 
higher multiplicities in an iterative manner, thereby accounting only 
for the leading logarithmic terms from parton-level processes with 
higher multiplicity. Depending on the 
jet multiplicity, higher perturbative 
orders correspond to different powers of the 
QCD coupling constant: the leading order prediction for $n$-jet production is 
proportional to $\alpha_s^{n-2}$. 
So far, fixed-order calculations were available 
up to next-to-next-to-leading order (NNLO) for two 
jets~\cite{babis2j,our2j,weinzierl2j}, up to next-to-leading order 
(NLO) for three~\cite{ERT,kunszt,event} and 
four jets~\cite{dixonsigner4j,nagy4j,cullen4j,weinzierl4j}. 
For five and more jets, only leading order calculations were 
available~\cite{tree5p,moretti6j,amegic}. For jets involving massive quarks,
NLO results are available for three-jet final states~\cite{quarkmass}.

Calculations based on parton showers, 
incorporated in multi-purpose event generator 
programs~\cite{herwig,ariadne,pythia},  
provide a satisfactory description of multi-jet production rates. Since 
these programs contain many tunable phenomenological parameters, their 
predictive power is however very limited. 

In this letter, we present the first calculation of NNLO corrections to 
three-jet production and the  next-to-next-to-next-to-leading order (N$^3$LO)
corrections to two-jet production in $e^+e^-$ annihilation. Together with the 
previously available NLO corrections to four-jet production and the 
leading-order description of five-jet final states, these are used for a fully
consistent perturbative description of  $e^+e^- \to$~jets at order $\alpha_s^3$
in perturbative QCD.

 The calculation of the $\alpha_s^3$ corrections for three-jet production
 is carried out using 
the newly developed
parton-level event generator program {\tt EERAD3} which contains 
the relevant 
matrix elements with up to five external partons~\cite{3jme,muw2,V4p,tree5p}. 
Besides explicit infrared divergences from the loop integrals, the 
four-parton and five-parton contributions yield infrared divergent 
contributions if one or two of the final state partons become collinear or 
soft. In order to extract these infrared divergences and combine them with 
the virtual corrections, the antenna subtraction method~\cite{cullen4j,ant} 
was extended to NNLO level~\cite{ourant} and implemented
for $e^+e^- \to 3\,\mathrm{jets}$ 
and related event-shape variables~\cite{eerad3} into {\tt EERAD3}. 
The analytical cancellation of all 
infrared divergences serves as a very strong check on the implementation. 

Initial results obtained with {\tt EERAD3} on NNLO corrections to  
event shape observables were reported in~\cite{ourevent} and 
applied in the extraction of the strong coupling constant from LEP
data in~\cite{ouras}. Since the program provides the full kinematical 
information for each event, it can also be used to 
simultaneously compute the production cross sections
 for three, four and five jets through to ${\cal O}(\alpha_s^3)$
for any infrared-safe jet algorithm and as function of the jet resolution 
parameter. The jet rates are then defined by normalizing the 
multi-jet cross sections to the total hadronic cross section
computed at the same order. 
\begin{figure}[t]
\begin{center}
\epsfig{file=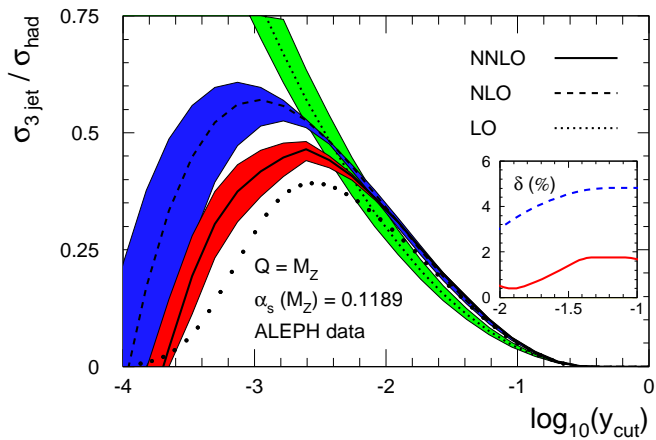,angle=-90,width=8.5cm}
\end{center}
\caption{Perturbative fixed-order description of the three-jet rate 
at $Q=M_Z$, compared to data obtained with the ALEPH 
experiment~\protect{\cite{aleph}}. 
Experimental errors are too small to 
be visible on the figure.\label{fig:3j}}
\end{figure}

The four-jet~\cite{dixonsigner4j,nagy4j,cullen4j,weinzierl4j} 
and five-jet rates~\cite{tree5p} were known previously to
${\cal O}(\alpha_s^3)$. Our major new result is the three-jet rate to 
this order, which corresponds to NNLO in the perturbative expansion. 
Figure~\ref{fig:3j} displays the three-jet rate 
at LEP1 energy $Q=M_Z$ as function of the jet 
resolution $\ycut$
 at LO, NLO, NNLO. At NNLO, the denominator has been expanded, 
as described in~\cite{ourevent} to contain only terms up to 
${\cal O}(\alpha_s^3)$ in the jet rate. 
The theoretical uncertainty band is
defined by varying the renormalization scale $\mu$
in the coupling constant in the interval $M_Z/2 < \mu < 2\,M_Z$, 
and the world average value~\cite{bethke} $\alpha_s(M_Z) = 0.1189$
 is  used, consistently evolved to other scales at each order.
The fixed-order theoretical predictions 
for three-jet rate become negative for small values of 
$\ycut$, where fixed order perturbation theory is not applicable due to the 
emergence of large logarithmic corrections at all orders, 
requiring resummation~\cite{durham,zanderighi}. 
We therefore restrict our comparison to $\ycut > 10^{-4}$, 
although data at lower jet resolution parameters are available.  
 
For large values of $\ycut$, $\ycut > 10^{-2}$, the NNLO corrections 
turn out to be very small, while they become substantial for medium and 
low values of $\ycut$. The maximum of the jet rate is shifted towards 
higher values of $\ycut$ compared to NLO, and is in better 
agreement with the 
experimental observation. 
The theoretical uncertainty is lowered considerably compared to NLO. 
Especially in the region $10^{-1}> \ycut > 10^{-2}$, which is relevant 
for precision phenomenology, one observes a reduction by almost a factor 
three, down to below two per cent relative uncertainty. 
 Since the error band in this region is barely visible in the plot, we 
display the relative theoretical uncertainty 
\begin{displaymath}
\delta = \frac{\max_{\mu} (\sigma(\mu)) - \min_\mu (\sigma(\mu)) }
{2 \sigma (\mu = M_Z)}
\end{displaymath}
at NLO and NNLO 
as an inset. The relative 
uncertainty on the LO calculation is constant at 10.2\%.

The fixed-order 
NNLO description is still above the
data at low jet resolution, where the convergence of the 
perturbative series is spoilt by large logarithms of $\ycut$ at all orders, 
and where a resummation should be carried out~\cite{durham}. Furthermore, 
the theoretical parton-level prediction is compared to hadron-level 
data, thereby neglecting hadronization corrections, which may also 
account for part of the discrepancy.

To compute the jet rates with different multiplicities, it is 
more appropriate to normalize all jet cross sections to the 
total hadronic cross section corrected 
to third order~\cite{kuhnrev}
 in the QCD coupling constant, ${\cal O}(\alpha_s^3)$. We consistently 
neglect numerically small QCD singlet contributions at this order, which 
were found to contribute at most one per cent~\cite{kuhnrev} to the total 
coefficient of the  ${\cal O}(\alpha_s^3)$ correction, and which are
equally small in the individual jet multiplicities~\cite{dixonsigner4j}.
The total hadronic cross section is made up from the 
sum over all jet multiplicities. At ${\cal O}(\alpha_s^3)$, this sum 
runs from two-jet through to five-jet final states, 
such that the corresponding jet rates must add to unity. Consequently, our 
calculation yields the N$^3$LO 
expression for $e^+e^- \to 2$~jets as a by-product. It is interesting to 
note that some earlier NNLO calculations of the two-jet 
rate~\cite{our2j,weinzierl2j} were essentially exploiting the same feature 
at ${\cal O}(\alpha_s^2)$.  
\begin{figure*}[t]
\begin{center}
\parbox{15.8cm}{\epsfig{file=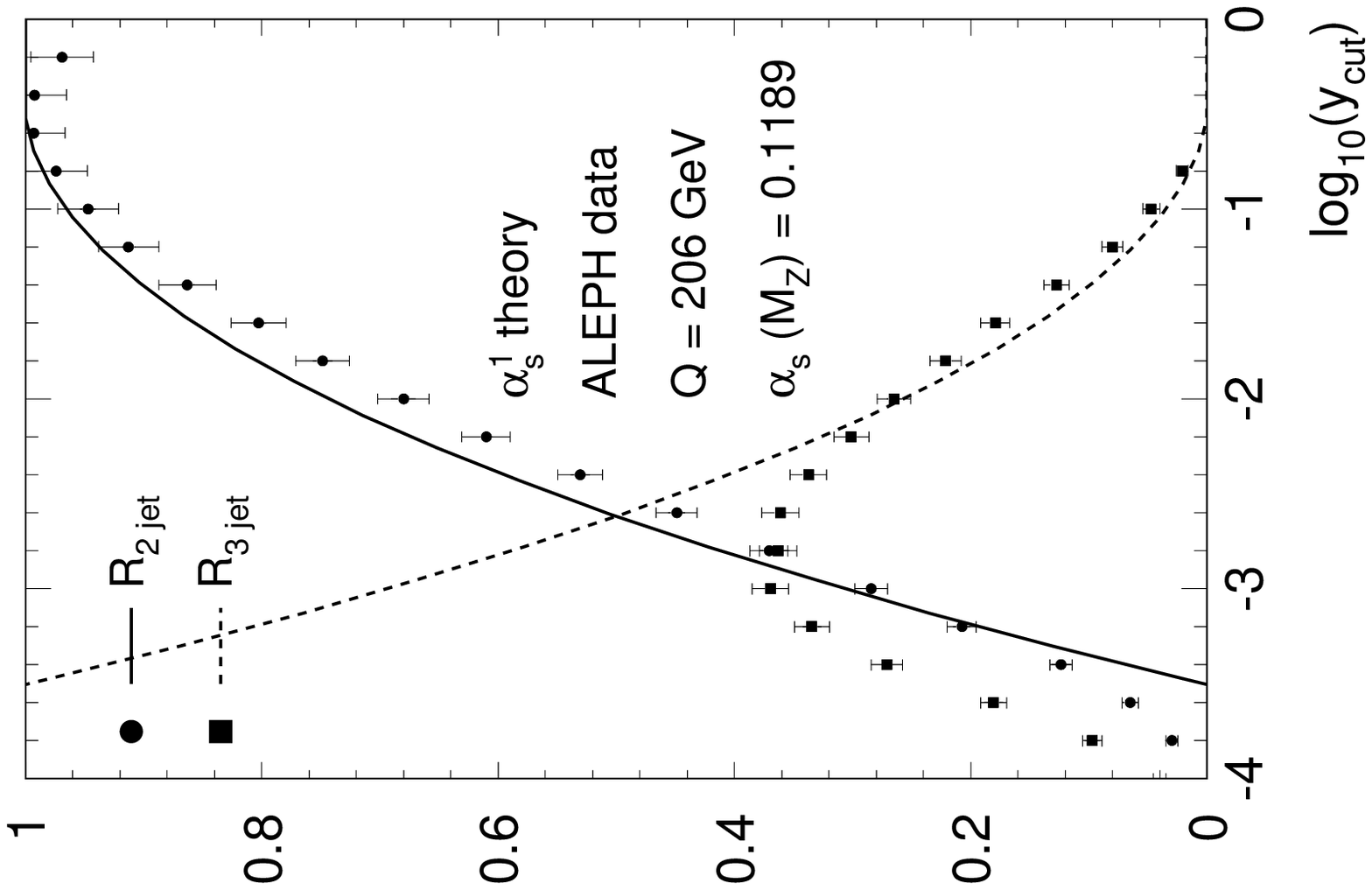,angle=-90,width=5.0cm}
\epsfig{file=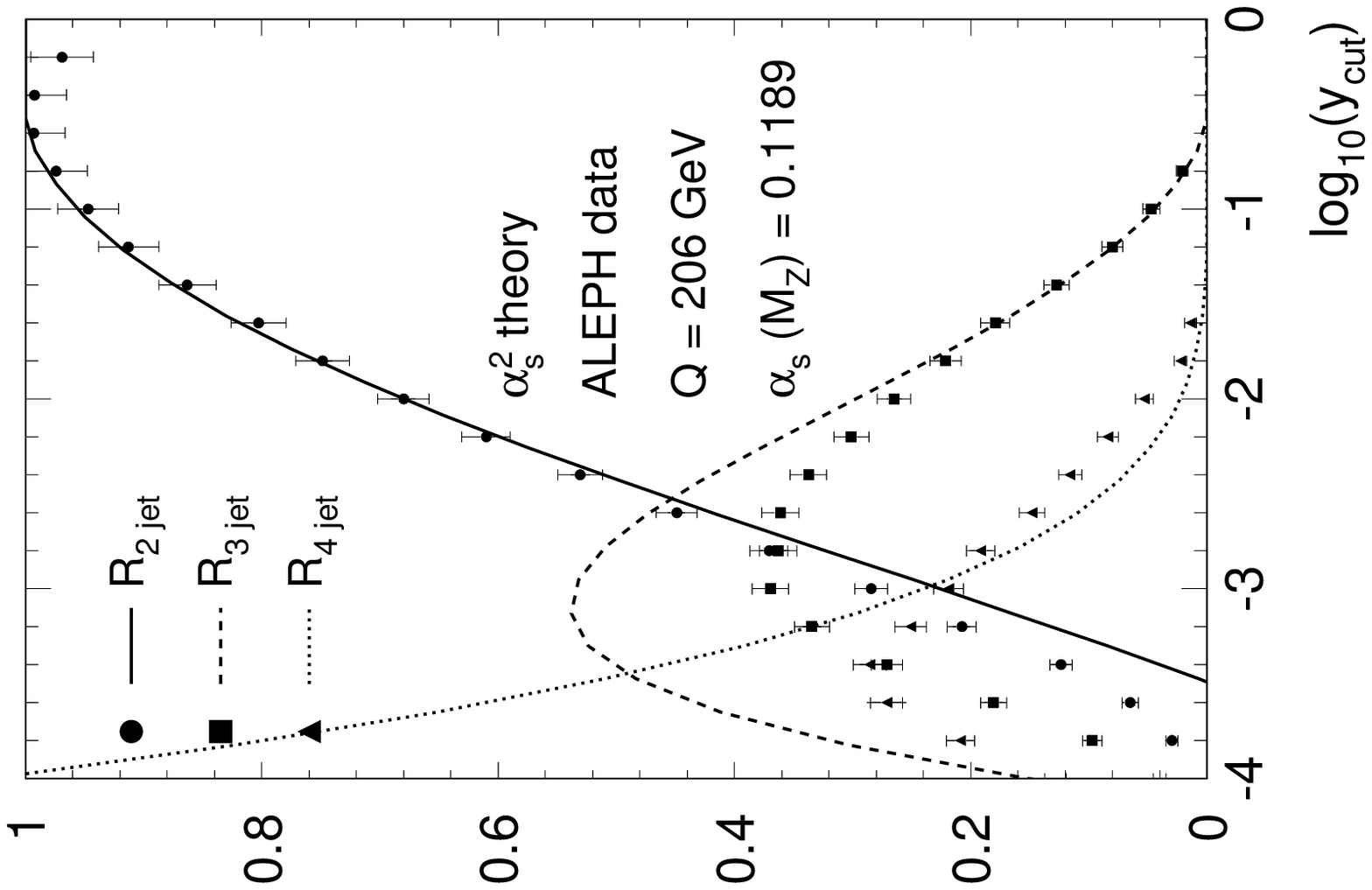,angle=-90,width=5.0cm}
\epsfig{file=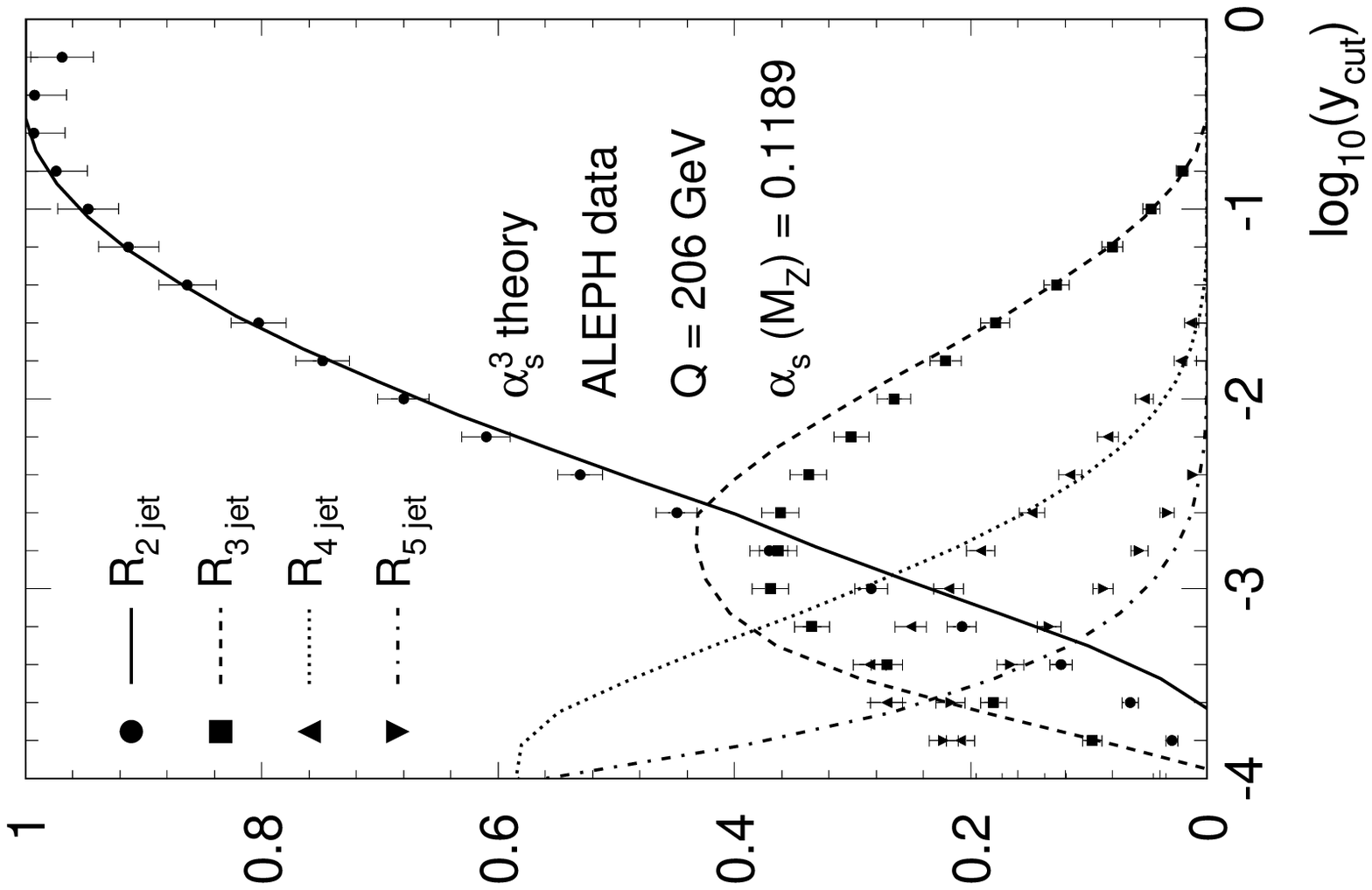,angle=-90,width=5.0cm}}
\end{center}
\caption{Jet rates at first, second and third order in the 
strong coupling constant, compared to data 
from ALEPH~\protect{\cite{aleph}}.
The rates are normalized to the total
hadronic cross section at that order.
\label{fig:rates}}
\end{figure*}

Figure~\ref{fig:rates} shows the parton-level theoretical predictions for the 
jet fractions at first, second and third
order
in the  strong coupling constant, compared to experimental hadron-level 
data from ALEPH~\cite{aleph}. 

By comparing the three plots, we observe that there is systematically improved
agreement for each of the jet rates as the order of perturbation theory
increases.  At each order a new multi-jet channel opens up, e.g. the five-jet
rate at ${\cal O}(\alpha_s^3)$, which is positive definite and essentially
monotonically increasing at small  $\ycut$.  Since all jet rates are normalized
to unity, the new five-jet channel has the effect of reducing the contribution
to the two-jet, three-jet and four-jet rates, in the region of $\log_{10}(y_{cut})$
where the five-jet rate contributes.  One very clear effect is to cause the
turnover in the four-jet rate (which is not present at ${\cal O}(\alpha_s^2)$). 
A second effect is to add more structure to the shape of the two- and three-jet
rates, which lie much closer to the data for $\log_{10}(\ycut) < -2.5$.    Of
course, the effect of the higher order corrections also extends to larger values
of $y_{cut}$, due to the different contributions of the two-loop virtual and
virtual-radiation graphs to the three- and four-jet rates, as well as the way
that the double radiation contribution interacts with the jet algorithm, and
through the normalization to the total hadronic cross section.   This is visibly
less dramatic, but by adding more structure to the theoretical prediction,
enables a better description of the data.

Previous experimental studies of multi-jet production rates compared only 
with standard leading-order parton shower event generator programs, which 
yielded a good description of the data at the expense of large hadronization 
corrections~\cite{reviews,aleph}. In the light of our new results, this 
issue should be carefully reexamined within fixed-order perturbation theory. 

In this letter, we reported on the NNLO QCD corrections to 
the three-jet production rate at parton-level 
in $e^+e^-$ annihilation, which is the 
first genuine NNLO 
calculation of a jet production rate at particle colliders. We observed that 
(hadron-level) experimental three-jet data are described considerably better 
in shape and normalization, and over a wider range in $\ycut$,  than 
at NLO. 

At the same order in the strong coupling constant, $\alpha_s^3$, we describe 
four-jet production at NLO and five-jet production at LO, reproducing 
earlier results. By combining those and normalizing to the total hadronic 
cross section at this order, we obtained the two-jet rate to N$^3$LO 
in perturbation theory as a by-product. We observe that with increasing 
order in the strong coupling constant, the multi-jet rates are 
better described over an increasing range of resolution parameters. Our 
results clearly highlight how perturbative QCD successfully describes 
jet production rates at the parton-level.

{\bf Acknowledgments:} 
Part of this work was carried out while the authors were attending 
the programme ``Advancing Collider Physics: From 
Twistors to Monte Carlos'' of the 
Galileo Galilei Institute for
Theoretical Physics (GGI) in Florence. We thank the GGI
 for its hospitality and the 
  Istituto Nazionale di Fisica Nucleare (INFN) for partial support.
This research was supported in part by the Swiss National Science Foundation
(SNF) under contract 200020-117602,  
 by the UK Science and Technology Facilities Council and  by the European Commission's Marie-Curie Research Training Network under contract
MRTN-CT-2006-035505 ``Tools and Precision Calculations for Physics Discoveries
at Colliders''.


\end{document}